\documentclass[12pt]{amsart}
\usepackage{geometry} 
\newtheorem{definition}{Definition}

\title{Superlogic Manifolds and Geometric approach to Quantum Logic }

\date{} 

\begin{document}

\maketitle

\begin{center}

\vspace{1cm}

{\bf Joseph Kouneiher}\footnote{e-mail: joseph.kouneiher@unice.fr}$^{, {\dag}}$
{\bf and Newton da Costa}\footnote{e-mail: ncacosta@terra.com.br}$^{,\star}$

\vspace{.5cm}

{\it
${\dag}$ESPE-Universit\'e de Nice Sophia Antipolis, 89, Av. GeorgesV,   \\
F-06064 Nice cedex 01 \\
\vspace{0.2 cm}
$^{\star}${Universidade Federal de Santa Catarina, Florian—polis, Brazil }
}
\end{center}

\begin{abstract}
\noindent The main purpose of this paper is to present a new approach to logic or what we will call superlogic. This approach constitutes a new way of looking at the connection between
quantum mechanics and logic. It is a {\it geometrisation} of the Quantum logic. Note that this superlogic  is not distributive reflecting a good propriety to describe quantum mechanics, non commutative spaces and contains a nilpotent element.
\end{abstract}

\section{Introduction}

\noindent                                                                                                                                                                                                                                                                                                                                                              \noindent In 1666, G.W. Leibniz envisaged a universal scientific language, the {\it characteristica universalis}, together with a symbolic calculus, the {\it calculus ratiocinator}, for formal logical deduction within this language. Leibniz soon turned his attention to other matters, including the creation of the calculus of infinitesimals, and only partially developed his logical calculus. Nearly two centuries later, in Mathematical Analysis of Logic (1847) and Laws of Thought (1854), G. Boole took the first decisive steps toward the realization of Leibniz's projected calculus of scientific reasoning.

\noindent The genesis of Quantum logic began with J. von Neumann in 1932 \cite{neuman1}. His main argument was  that certain linear operators, the projections defined on a Hilbert space, could be regarded as representing experimental propositions affiliated with the properties of a quantum mechanical system. He wrote,

\begin{quote}
{\it  "...the relation between the properties of a physical system on the one hand, and the projections on the other, makes possible a sort of logical calculus with these."}
\end{quote}

\noindent Later on, in1936, von Neumann published with G. Birkhoff a definitive article on the logic of quantum mechanics  \cite{birkhoff1, birkhoff2}. In this paper, Birkhoff and von Neumann proposed that the specific Quantum logic of projection operators on a Hilbert space should be replaced by a general class of Quantum logics governed by a set of axioms, much in the same way that Boolean algebras had already been characterized axiomatically. They observed that, for propositions $P, Q, R$ pertaining to a classical mechanical system, the distributive law

     $$       P \wedge (Q \vee R) = (P \wedge Q) \vee (P \wedge R) $$

\noindent holds, they gave an example to show that this law can fail for propositions affiliated with a quantum mechanical system, and they concluded that,

\begin{quote}
{\it   "...whereas logicians have usually assumed that properties of negation were the ones least able to withstand a critical analysis, the study of mechanics points to the distributive identities as the weakest link in the algebra of logic."}
\end{quote}

\noindent Birkhoff and von Neumann went on to argue that a Quantum logic ought to satisfy only a weakened version of the distributive law called the modular law; however, they pointed out that projection operators on a Hilbert space can fail to satisfy even this attenuated version of distributivity. Much of von Neumann's subsequent work on continuous geometries \cite{neuman2} and rings of operators \cite{neuman3} was motivated by his desire to construct logical calculi satisfying the modular law. In 1937, K. Husimi \cite{husimi} discovered that projection operators on a Hilbert space satisfy a weakened version of the modular law, now called the orthomodular identity.

\noindent The other interesting breakthrough was taking in 1957 by G. Mackey, who  wrote an expository article on quantum mechanics \cite{mackey1} based on lectures he was giving at Harvard. In 1963, he published an expanded version of these lectures in the form of an influential monograph \cite{mackey2}. Note that Mackey's questions form an orthomodular lattice. The simplicity and elegance of Mackey's formulation and the natural and compelling way in which it gave rise to a system of experimental propositions inspired a renewed interest in the study of Quantum logic, now identified with the study of orthomodular lattices. 

\noindent In 1964, C. Piron introduced an alternative to Mackey's approach in which questions again band together to form an orthomodular lattice, but this time possessing more of the special features of the lattice of projection operators on a Hilbert space \cite{piron}. 

\noindent A list of more or less "natural conditions" on generalized Hilbert spaces was soon proposed in the hopes of singling out the "true" Hilbert spaces. In 1980, H. Keller dashed these hopes by constructing an example of a generalized Hilbert space satisfying all of the proposed natural conditions, but that is not a standard Hilbert space \cite{keller}. 

\noindent Even more, in orthodox quantum mechanics, the combined system (when systems are combined or coupled to form composite systems) is represented mathematically by a so-called tensor product of Hilbert spaces. Very early many researchers realized that the entire Quantum logic program would falter unless a suitable version of tensor product could be found for the more general logical structures then under consideration.

\noindent Composite physical systems were studied from the perspective of Quantum logic in an important and influential sequence of papers by D. Aerts \cite{aerts}. In parallel with the development of Quantum logic, and starting as early as 1970 \cite{davies}, Davies, Lewis, Holevo, Ludwig, Prugovecki, Ali, Busch, Lahti, Mittelstaedt, Schroeck, 'Bujagski, Beltrametti \cite{beltrametti}, et al worked out a theory of quantum statistics and quantum measurement based on so-called effect operators on a Hilbert space. Every projection operator is an effect operator, but not conversely, and the effect operators do not even form a lattice, let alone an orthomodular lattice, or even an orthoalgebra.

\noindent In fact we can consider that during the years 1938-1994 Quantum logic has been under development for roughly half a century. The history of Quantum logic has been a story of more and more general mathematical structures - Boolean algebras, orthomodular lattices, orthomodular posets, orthoalgebras, and effect algebras - being proposed as basic models for the logics affiliated with physical systems. \\

\noindent  In the rest of the paper we will present a new  approach to the question of Quantum logic. The main idea is to introduce the correlation or interaction already at the level of the elements of the logic or superlogic.  Thus this paper presents a kind of  '{\it geometrization}' of  Quantum logic\footnote{ Or {\it 2-dimensional logic}. Indeed,   as in the case of real number where ${\mathbb R}$ is considered as one dimension, ${\mathbb C}$ as 2-dimensions $(\mathbb{R}, \mathbb {R})$ with the structure $i^2 = -1$ and  ${\mathbb Q}$ (quaternion) as 4 dimensions with three $i, j, k$ complexes structures. So we can think of the superlogic as a couple of classical logic with a structure $n^2 = 0$, i.e. as 2-dimensional logic. It is even more relevant to think this superlogic as in the case of supersymmetry where we have a commuting and anticommuting (or Grassmann) coordinateds.} that could be extended to all logic\footnote{See \cite{kouneiher} for introduction to the geometrization of classical logic and the propositional manifolds}. This is fulfilled through the  introduction of a new structure $n$ such that $n^2 =0$ and sheaves of fields on the superlogic manifolds. So it codes the non commutativity   and  non distributivity of the quantum formalism\footnote{Moreover, in some manner our superlogic is  linked with a generalized version of {\it Bohr Topos} or more precisely a certain presheaf of Bohr toposes. Indeed, as we know, one might think of a Bohr topos as (part of) a formalization of the ÒcoordinationÓ of the physical theory of quantum mechanics, providing a formalized prescription of how to map the theory to propositions about (experimental) observables of the system, i.e.  a `topos-theoretic formulation of physics".  However, Bohr toposes currently formalize but one aspect of quantum mechanics, namely ``{\it the quantum mechanical phase space}" in the form of the quantum observables and the quantum states. The plain Bohr topos does not encode any dynamics, though in the spirit of $AQFT$ a certain presheaf of Bohr toposes on spacetime does encode dynamics \cite{jeremy, doring}.}.

\noindent  Logically, the novelty of this approach consists in a new way of treating propositions. It is inspired by the analogy between propositions and measurements in physics, in particular
quantum measurements. Suppose a physicist performs a measurement of a
certain physical quantity $A$ pertaining to a certain physical system. Suppose
this system is in a certain state $x$. The physicist will then in general formulate
the result of his measurement as a proposition of the form $A = \mu$, where $\mu$ is
the value of $A$ measured, and he will then claim this proposition to be a true
statement about the system under investigation. In this there is no difference
between classical and quantum physics. There exists, however, an essential
difference between the classical and the quantum case which seems to us of
fundamental importance from the logical point of view. Namely, the meanings
of the physicist's assertion that $A = \mu$ is true differ in the two cases, classical
and quantum. In classical mechanics the proposition $A = \mu$ is a true statement
about the physical system in state $x$. In the quantum case it is a true statement
too. The crucial difference, however, is that in the quantum case the proposition
$A = \mu$ is in general no longer a true statement about the state $x$ but about a
certain state $y$ distinct from $x$, namely about the state of the system `after
measurement". The reason for this is that quantum measurements generally
involve, in contrast to 'classical' measurements, a change of state of the system
measured. Logically speaking, the situation we have in classical mechanics is
this. Given a state $x$ (state of affairs, state of the world...) and some proposition
$\alpha$. Then $\alpha$ has some truth value in state $x$. In bivalent logic these truth values
are 'true' and 'false'. In multi-valued logic there are more truth values, possibly
even infinitely many. The situation in quantum mechanics is different. Given
a state $x$ and a proposition $\alpha$. Then $\alpha$ does not necessarily possess any truth
value in $x$. Rather it is only in some other state distinct from $x$ , namely in
the state 'after measurement', that it acquires a truth value.

\section{Definition of Superlogic}

\noindent  Superlogic manifolds\footnote{They are the analogous to Supermanifolds which are special cases of noncommutative manifolds, the local structure of supermanifolds makes them better suited to study with the tools of standard differential geometry and locally ringed spaces.} have a structure analogue to $\mathbb{ R}^{m|\mid n}$ which is $\mathbb{Z}/2$-graded vector spaces with $\mathbb{R}^m$ as the even subspace and $\mathbb{R}^n$ as the odd subspace or Grassmann's numbers. We will use the notation $\mathbb{ L}^{m|\mid n}$. This superlogic codes the non commutativity and non distributivity of the Quantum mechanics theories and non commutative spaces. 

\noindent We start by defining the language of super propositional logic which is built up from the following symbols.  As we said our superlogic has the $ \mathbb{Z}/{2\mathbb{Z}}  [n] $ structure.

\noindent  The elements of this superlogic are the couples $(P,Q)$ with the structure $n$ where $n^2=0$. The elements of $\mathbb L$ can be writting as $ L= P+ n Q $.The property of  $n^2=0$ means that all we can have are terms of at most degree $1$ in $n$.

\noindent Some remarks are in order. Here, we should think of the symbol $ '+'$ as in the case of complex (and quaternion numbers) where $z \in {\mathbb C}$ can be writing as $z = x +iy$. It means that each point of our supermanifolds is described by a couple of coordinates or said differently composed from two entities, the first one is a commuting coordinate (or propositions) and the second one an anti commuting coordinate. Now, by commuting coordinates we should understand that the order in this case is not an issue, i.e. $P_iP_j = P_jP_i$ as in the classical case.  Regarding the anti commuting coordinates (or propositions), here we should be careful about the order, as in the case of the commutators in quantum mechanics and quantum fields theories. More precisely $Q_iQ_j = - Q_jQ_i$. For instance, in quantum mechanics the position and momentum would be described by an anti commuting coordinates in our supermanifolds. For instance, this can help us to describe, from logical point of view, the non abelian gauge theories or/and theories unifying the space-time and non abelian gauge formalism and their quantization.

 \noindent  Let $P$ and $Q$  be an element of propositional logic or predicate logic, this superlogic can generate the others logical systems by differentiation, Note that this time we have four values of truth $ 0,1,n,1+n $. As an illustration we can think of $P$ as property of an object or probability to have such property and $nQ$ the probability to follow some path. As we will see this property will be useful to describe interference in quantum mechanics.

\section{The rules}

\begin{equation}
\neg L = \neg P + n \neg Q
\end{equation} 

\noindent Regarding the negation operation it should be understood as map which associate to each proposition $P$ its opposite $\neg P$. Sometimes, as in the case of measurement of the spin for instance, we can instantiate the negation map in terms of symmetry: measuring the Spin $S$ or $-S$.

\begin{eqnarray}
L \wedge L' & = & (P +nQ) \wedge (P' + nQ')  \\  \nonumber
& = & P \wedge P' + n P \wedge Q' + n Q\wedge P' + nQ \wedge nQ' \\ \nonumber
& = & P \wedge P' + n P \wedge Q' + n Q\wedge P'\\ \nonumber
& = & P \wedge P' + n (P \wedge Q' + Q\wedge P')
\end{eqnarray}

because $n^2=0$.

\begin{eqnarray}
L \vee L' & = & (P +nQ) \vee (P' + nQ')  \\  \nonumber
& = & P \vee P' + n P \vee Q' + n Q\vee P' + nQ \vee nQ' \\ \nonumber
& = & P \vee P' + n P \vee Q' + n Q\vee P'\\ \nonumber
& = & P \vee P' + n (P \vee Q' + Q\vee P')
\end{eqnarray}
because $n^2=0$. We will call the second term : the superposition term, it is an essential difference with the other logic. Indeed the superposition terms it come naturally

Distributivity I
\begin{eqnarray}
\nonumber
L\vee (L' \wedge L") & = & (P+ nQ) \vee [(P' + n Q') \wedge (P" + n Q")] \\ \nonumber
& = & (P + nQ) \vee [(P' \wedge P") + (P' \wedge n Q") \\
& + & (nQ' \wedge P") + (nQ' \wedge nQ")]\\ \nonumber
& = & (P + nQ) \vee [(P' \wedge P") + (P' \wedge n Q") + (nQ' \wedge P") ] \\ \nonumber
& = & (P \vee (P' \wedge P")) + (P \vee (P' \wedge n Q")) + (P \vee (nQ' \wedge P")) \\ \nonumber
& + & (nQ \vee (P' \wedge P")) + (nQ \vee (P' \wedge n Q")) + (nQ \vee (nQ' \wedge P")) \\ \nonumber
& = &  (P \vee (P' \wedge P")) +  (P \vee (P' \wedge n Q")) \\ \nonumber
& + & (P \vee (nQ' \wedge P")) + (nQ \vee (P' \wedge P")) 
\end{eqnarray}

Distributivity II
\begin{eqnarray}
(L\vee L'  )\wedge L" & = & (P\vee P')\wedge P" + (P\vee P') \wedge n Q" \\ \nonumber
& + & (P\vee nQ') \wedge P" + n(Q\vee P') \wedge P"
\end{eqnarray}

in case $n =0$, we found the old equality : 

$$  L\vee (L' \wedge L")  =  (P \vee (P' \wedge P")) $$

and 

$$ (L\vee L' ) \wedge L"  =   (P\vee P')\wedge P"$$

 \section{Propositional (or Predicates) Supermanifolds}

 Let $P_i$ and $Q_j$, ($i,j \in I \times J$, $I$ and $J$ are set of indexes) be two sets of open Boole algebras of propositions (or predicates), a propositional supermanifold is defined as following:

 \begin{definition}
 $\forall (i, j) \in (I\times J), \,\, I\times J \rightarrow \{0,1\}^{P_i}\times \{0,1\}^{P_j},$
 \end{definition}
 where $\{0,1\}^{P_i}$ ($\{0,1\}^{P_j}$) denote the maps of $P_i$ ($P_j$) into $\{0,1\}^{P_i}$  ($\{0,1\}^{P_j}$)(see \cite{kouneiher} for more details and definition of transitions functions, coccyle conditions and for the equivalence relation).
 
 \begin{definition}
 we provide the Boolean supermanifold $\mathbb L$ with the following operations:
 $$\neg : {\mathbb L} \rightarrow {\mathbb L} $$
$$\vee : {\mathbb L}^2 \rightarrow {\mathbb L} $$
$$\wedge : {\mathbb L}^2 \rightarrow {\mathbb L} $$
$$\Delta : {\mathbb L} \rightarrow {\mathbb L}^2 $$
$$\vdash : {\mathbb L} \rightarrow \{0,1\} $$
$$n : {\mathbb L} \rightarrow n{\mathbb L}, \,\,\, with \,\,\, n^2=0 $$
\end{definition}

\section{Vectors fields}

\begin{definition} 
A vector field overs a   supermanifolds of propositions $\mathbb L$ is a continuous application $X$ :
$$ X(L) = X(P + nQ) = X(P) + nX(Q),$$
$$ X(\neg L) = \neg X(L),$$
$$ X(L \vee K) = [X(L) \vee K] \wedge [L \vee X(K)],$$
$$ X (L \wedge K) = [ X(L) \wedge K ] \vee [L \wedge X(K)] $$
\end{definition}

\noindent where $\vee$ is 'OR' and $\wedge$ is "AND'.\\

\noindent  Such a set of vector admit the following operations :
\[
[\neg X] (L) = \neg [X(L)] 
\]
\[
[X \vee Y] (L) = X(L) \vee Y(L) 
\]
\[
[X \wedge Y](L) = X(L) \wedge Y(L).
\]

\section{Supefields space}

\noindent One may define functions from this vector space to itself, which are called superfields. The above algebraic relations imply that, if we expand our superfield as a power series in $n$ and  then we will only find terms at the zeroeth and first orders, because $n^2 =  0$. Therefore superfields may be written as arbitrary functions of $P$ multiplied by the zeroeth and first order terms in  Grassmann coordinates

           $$ \Phi (L) =\Phi(P)+ nQ \Phi(P) + \Phi(nQ)$$

\noindent  If $\mathcal M$ is a supermanifold of dimension $(k,l)$, then the underlying space $M$ inherits the structure of a differentiable manifold whose sheaf of smooth functions is $O_{\mathcal M}/I $, where $I$ is the ideal generated by all odd functions. Thus $M$ is called the underlying space, or the body, of $\mathcal M$. The quotient map $O_{\mathcal M} \rightarrow O_{\mathcal M}/I $ corresponds to an injective map $ M \rightarrow \mathcal M$; thus $M$ is a submanifold of $\mathcal M$.

\section{Superlogical Cohomology}
\begin{definition}
A character for a set of propositions $\mathbb L$ is a function $\chi$ with values in $\{0,1\}$ such that :
\[ \chi(L) = \chi(P + nQ) = \chi(P) + \chi(nQ) \]
\[
\chi(\neg L) = \chi^{op}(L),\]
\[ \chi (L \vee K) = \chi(L) + \chi(K) \]
\[\chi(L\wedge K) = \chi (L)\chi(K) \]
\[ [L \implies K] \implies [\chi(L) \leq \chi(K)] \]
\end{definition}

\noindent Let $\mathcal X$ be the set of all characters. We have the following diagrams:

\[ \Delta :  {\mathcal X} \rightarrow {\mathcal X} \otimes {\mathcal X}, \, \, \, x \otimes x \]
\[ \epsilon : {\mathcal X} \rightarrow {\bf F}_2 \mapsto 1, \]
\[ \mu : {\mathcal X} \otimes {\mathcal X} \rightarrow {\mathcal F}, \]

$\mathcal F$, are the functions over $\mathbb L$ with values in ${\bf F}_2$, with ${\bf F}_2 = {\mathbb Z}/2{\mathbb Z}$. 

\begin{definition}
Let $\mathcal A$ denote the algebra generated by the characters of $\mathbb L$.
The Cohomology of $\mathbb L$ is : $H^*(\Omega ({\mathcal A}, {\bf F}_2)$, where ${\bf F}_2 = {\mathbb Z}/2{\mathbb Z}$. 
\end{definition}

\noindent An essential property is the compatibility with the exact sequences of Boolean manifolds.

\section{Applications to Quantum Mechanics}

\noindent Suppose we prepare a set of electrons to pass through the two-slit experiment. Part of those electrons are prepared to pass through the first hole  and the others through the second hole. But quantum mechanics teach us that the probability of the first part to pass through the second hole is not zero, and vice versa for the second part so we have the situation:\\

\noindent $(P_1+ n Q_{12})$ is the probability $P_1$ that the part I of the electrons pass through the first hole with the probability $nQ_{12}$ to pass through the second hole. \\

\noindent $(P_2 +n Q_{21})$ is the probability $P_2$ that the second part of the electrons pass through the second hole with the probability $nQ_{21}$ to pass through the first hole.\\

\noindent  So when we send the set altogether, we'll have

$$(P_1 +nQ_{12}) \wedge (P_2 +nQ_{21}) = P_1\wedge P_2 + P_1 \wedge nQ_{21} +  nQ_{12} \wedge P_2 + (terms\,\, order\,\, n^2 = 0)$$

\noindent So our Superlogic, describe exactly the two-slit experience:

\noindent The first term  $P_1 \wedge P_2$ describe the probability that the first part of the electrons pass through the first hole and the second part of the electrons pass through the second hole, the second term $( P_1\wedge nQ_{21} +  nQ _{12}\wedge P_2)$,  the interference terms, describe the superposition of the two parts.  So here we get the result from the first order without the  need to a density of probability. Our Superlogic describe the two-slit experiment naturally. This same approach can be applied to the EPR experience to describe entanglement.

\noindent We can consider the second part  (or Grassmaniann coordinate) $(nQ)$ as term which describe the correlation between the objects of our experience, more precisely, it is a way to introduce the possibility of interaction already at the probability level.

\section{conclusion}

\noindent  Modern logic studies logical systems as formal systems based on a precisely defined formal language. The concept most
central to logic is that of logical consequence. Logical consequence is a relation
between two statements $\alpha$  and $\beta$  or, more generally, a relation between a set of
statements $\Sigma$ and a statement $\alpha$ . One may synonymously say "$\alpha$  is a logical
consequence of $\Sigma$" or "$\alpha$  follows (logically) from $\Sigma$" or "$\alpha$  can be deduced logically (or is deducible) from $\Sigma$". Logical deduction is a vital part of our
competence as human beings in both everyday and scientific discourse, and it is
one of the seminal achievements of modern mathematical logic to have provided
the tools for a mathematically rigorous analysis of the intuitive concept of logical.

\noindent  In this paper we present an axiomatization which give up the distributivity; and allow to combine the individual phenomena and the interaction one. A probability which take infinitesimal values (positive, but smaller than every positive real number); that allow probabilities to be imprecise (interval-valued, or more generally represented with sets of numerical values). It is a extension of the standard logic by introduce a structure coding the non commutativity of certains spaces and theories.

\end{document}